\documentclass[aps,prb,twocolumn,superscriptaddress]{revtex4-1}
\usepackage{graphicx}
\usepackage{picture}
\usepackage{xcolor}
\usepackage{color}
\usepackage{tikz}
\usepackage{ulem}
\usepackage{amsmath}
\usepackage{rotating}

\begin{document}
\title{Tuning Spectral Properties of Individual and Multiple Quantum Emitters in Noisy Environments}

\author{Herbert F Fotso}
\affiliation{Department of Physics, University at Albany SUNY, Albany, New York 12222, USA}
\affiliation{Department of Physics, University at Buffalo SUNY, Buffalo, New York 14260, USA
}


\begin{abstract}

A quantum emitter in a dynamic environment may have its energy levels drift uncontrollably in time with the fluctuating bath. This can result in an emission/absorption spectrum that is spread over a broad range of frequencies and presents a challenging hurdle for various applications. We consider a quantum emitter in an environment that alters the energy levels so that the emission frequency is represented by a Gaussian random distribution around a given mean value with given standard deviation and correlation time. We study the emission spectrum of this system when it is placed under the influence of a periodic sequence of finite width $\pi$ pulses. We show that this external field protocol can effectively overcome spectral diffusion in this system by refocusing the bulk of the emission spectrum onto the pulse carrier frequency. We further consider two such emitters in different noisy environments and find that the two-photon interference operation can be made efficient by the sequence of finite width pulses applied on both systems. Finally, we show that an ensemble of nominally similar emitters, each with its different environment, and thus randomly shifted emission frequency, can have its overall emission spectrum that would otherwise be inhomogeneously broadened according to the random distribution, refocused onto a lineshape with a well-defined central peak that has the linewidth of an individual isolated non-noisy emitter. These results demonstrate for this specific model of noisy environments, the protection of spectral properties by an external control protocol here represented by a periodic sequence of finite width pulses.      

\end{abstract}

\maketitle

\section{Introduction}
\label{sec:introduction} 
The ability to control and protect from environmental variations the spectral properties of a quantum emitter in a dynamic environment is of significant importance for numerous applications extending from spectroscopy to a variety of fundamental operations of quantum information processing (QIP). Indeed, spectral diffusion, the random drift of the emission frequency of a quantum emitter with time \cite{AmbroseMoerner_spectralDiffusion_Nature1991, KMFuBeausoleil_PRL2009, PfaffHanson_Science2014, LyasotaKapon_SciRep2019, GaoHolmes_APL2019, VuralMichler_PRB2020, VuralMichler_PRB2020}, reduces the efficiency of essential QIP procedures such as two-photon interference, entanglement generation between distant quantum nodes, and coupling to cavities\cite{KambsBecher_NJP2018, AwschalomHansonZhou_opticsQIP, AtatureEnglund_Wrachtrup_NatRevMat_2018, HansonAwschalom_QIP_ss_08, NV_Review_PhysRep2013, CarterGammonQDcavity, JeffKimble_qtmInternet, NielsenChuangBook}. These operations typically require well behaved spectral signatures of the involved quantum emitters. For this reason, spectral features dominated by random fluctuations, are significant obstacles to the scalability of photon-mediated operations or QIP interfaces.
\cite{Gao_Imamoglu_NatComm2013, Kuhlmann_Warburton_QDOT_NatPhys2013, SantoriVuckovicYamamoto_QDOT_Nat2002, Humphreys_Hanson_NV_entanglement, Hanson_loopholeFree_Nature2015}

One avenue to help overcome these fluctuations is the design of increasingly pristine systems which is an onerous task. Although this approach is indeed important to reduce unwanted defects and randomness, it remains limited in its effectiveness given that some minimal fluctuations will likely persist in most realizations of solid state systems for instance. For this reason solutions based on external control field can play a unique role in addressing this problem \cite{Schroeder_Englund_SPE_NatComm2017, QDOTs_HOM_Atature, Aharonovich_Englund_Toth_SS_SPE, KMFuBeausoleil_PRL2009, Acosta_Beausoleil_PRL2012, Dreau_Jacques, Pfaff_Hanson_Science2014, Basset_Awschalom_PRL2011, FaraonEtalNatPhot, Hansom_Atature_APL2014, Crooker_Bayer_PRL2010, Calajo_Passante_PRA2017, JSLee_Khitrin_JPhysB2008, IDS_2017, LukinVuckovic_npj2020, LiuKumano_OptExpress2017}.  

In earlier studies, we examined the emission and absorption spectrum of two-level systems when they are driven by a variety of pulse sequences including a periodic sequence of $\pi_x$ pulses~\cite{FotsoEtal_PRL2016, FotsoDobrovitski_Absorption, Fotso_JPhysB2019}. We showed that for an emitter with emission frequency $\omega$, 
the emission spectrum could be made mostly independent of the constant detuning $\Delta$ with respect to the pulse carrier frequency $\omega_0$. We also showed that the Hong-Ou-Mandel (HOM) two-photon interference (TPI)~\cite{HOM} could have its efficiency enhanced for two distant emitters with different respective emission frequencies when they are both driven by the same periodic sequence of instantaneous $\pi_x$ pulses~\cite{Fotso_TPI_PRB_2019}.

In the present paper, we examine the situation of emitters in explicitly noisy environments. In particular, we study the emission spectrum of a TLS with detuning $\Delta(t)$ with respect to a reference frequency $\omega_0$ such that $\Delta(t)$ follows a random Gaussian distribution with standard deviation $\sigma_{\Delta}$, mean value $\Delta_0$ and correlation time $\tau_c$. We show that for a periodic sequence of finite width $\pi$-pulses well away from the ideal pulse limit, the emission spectrum of the noisy quantum emitter can be made minimally dependent on the noisy environment with the bulk of the emission spectrum occurring at the pulse carrier frequency and satellite peaks at $\pm \pi/\tau$ similar to the static detuning case. Next, we demonstrate enhancement of the HOM-type two-photon interference experiment for two explicitly noisy emitters in different environments such that their detunings are $\Delta_1(t)$ and $\Delta_2(t)$ with mean values $\Delta_{01}$ and $\Delta_{02}$, standard deviations  $\sigma_{\Delta 1}$ and $\sigma_{\Delta 2}$, correlation times $\tau_{c1}$ and $\tau_{c2}$. Finally, we show that for a dilute ensemble of emitters with randomly distributed emission frequencies among individual emitters, the emission spectrum that would otherwise be inhomgeneously broadened, can be refocused by the periodic sequence of finite width pulses.

The rest of the paper is organized as follows. In section \ref{sec:model}, we discuss the model for the emitter in a dynamic environment, describing the Hamiltonian and the master equation for the density matrix operator of the two-level system in the radiation bath under the influence of the control field. In section \ref{sec:methods}, we describe the methods that are used to obtain the emission spectrum for an individual pulse-driven noisy emitter subject to spectral diffusion, to characterize the two-photon interference operation between two such emitters and, finally, to obtain the emission spectrum of a dilute inhomogeneous ensemble of two-level systems under the influence of the control protocol. In section \ref{sec:results}, we present the results for the different situations discussed above before finishing with our conclusions in section \ref{sec:conclusion}.

\section{Model, Emitter in dynamic environment}
\label{sec:model}

\begin{figure}[htbp] 
\begin{center}
\includegraphics*[width=7.0cm, height=7.50cm]{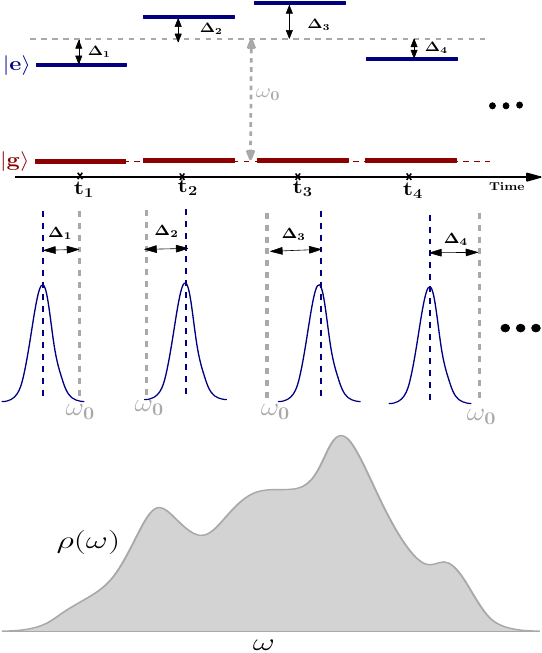}
    \caption{Schematic representation of spectral diffusion for a quantum emitter in a dynamic environment. The energy levels for the ground and excited states, $|g\rangle$ and $|e\rangle$ respectively, drift randomly in time around a target frequency $\omega_0$. Thus, at different points in time, the detuning of the emission with respect to $\omega_0$ takes random values $\Delta_1$, $\Delta_2$, $\Delta_3$, $\Delta_4$... As a result, the emission spectrum of the system over time adds up to a broad ill-defined lineshape.} 
\label{fig:schematic1}
\end{center}
\end{figure}

\begin{figure}[htbp] 
\begin{center}
\includegraphics*[width=7.0cm]{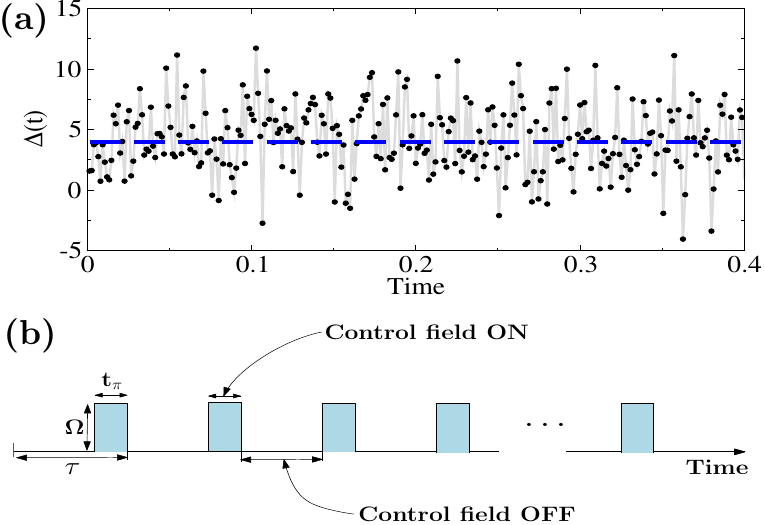}
\caption{(a): We study the emission spectrum of a TLS for which the emission spectrum with respect to a target frequency $\omega_0$ has a detuning $\Delta(t)$ that fluctuates in time following a random Gaussian distribution pictured here with $\Delta_0 = 4.0$ (dashed blue line), $\sigma_{\Delta} = 4.0$ and correlation time $\tau_c = 0.03$. (b): We will examine the spectral properties of the system under the influence of a periodic sequence of finite width $\pi$-pulses with inter-pulse delay $\tau$ and Rabbi frequency $\Omega$. The driving field is applied for a time $t_{\pi}$ that amounts to a $\pi_x$ rotation.} 
\label{fig:schematic2}
\end{center}
\end{figure}

We consider a quantum emitter represented by a two-level system (TLS). Its ground state $|g\rangle$ and its excited state $|e\rangle$ are separated by an energy $E_e - E_g = \hbar\omega_1 = \hbar(\omega_0 +\Delta)$. $\Delta$ is the detuning with respect to a target frequency $\omega_0$. In what follows, we set $\hbar = 1$. Because of the fluctuations in the environment, this detuning can vary randomly in time. Here, we will specifically consider the situation in which the fluctuations lead to a time-dependent detuning that follows a random Gaussian distribution centered around an average value $\Delta_0$, that has a standard deviation $\sigma_{\Delta}$ and a correlation time $\tau_c$. The TLS is coupled to a bosonic bath representing the normal modes of the radiation field. The protocol of interest in the present studies is represented by pulses at the target frequency $\omega_0$ with Rabbi frequency $\Omega_x(t)$ that is timed so as to impart on the emitter an appropriate $\pi$ rotation over a finite time $t_{\pi}$ before being switched off, allowing the system to evolve freely for a subsequent time $\tau - t_{\pi}$. The process is repeated periodically so that the entire sequence has period $\tau$. FIG.\ref{fig:schematic1} illustrates schematically the random drift in time of the emission frequency of a two level system: spectral diffusion. As a function of time, the detuning takes different random values $\Delta_1$, $\Delta_2$, $\Delta_3$, $\Delta_4$... As a results, the emission spectrum can be broadly ill-defined. FIG.\ref{fig:schematic2} shows a sampling of the detuning as a function of time for a random Gaussian distribution with average value $\Delta_0 = 4.0$, standard deviation $\sigma_{\Delta} = 4.0$ and correlation time $\tau_c = 0.03$  (a); and an illustration of the sequence of finite width pulses (b).

In the rotating wave approximation (RWA) and in the rotating frame, so that all energies are measured with respect to the target frequency $\omega_0$, the Hamiltonian describing this system can be written as:
\begin{eqnarray}
H = \sum_{k} \omega_k a^{\dagger}_{k}a_{k} &+& \frac{\Delta(t)}{2} \sigma_z - i \sum_{k} g_{k} \left( a^{\dagger}_{k} \sigma_- - a_{k} \sigma_+ \right) \nonumber \\
&+& \frac{\Omega_x(t)}{2}(\sigma_+ + \sigma_-). 
\label{eq:hamiltonian_1}
\end{eqnarray}

The operators $\sigma_z=|e\rangle\langle e|-|g\rangle\langle g|$, $\sigma_+ =|e\rangle\langle g|$, and $\sigma_- = |g\rangle\langle e| = (\sigma_+)^\dagger$ are respectively, the $z$-axis Pauli matrix, the raising, and the lowering operators for the two-level system. $a_k$ ($a^{\dagger}_k$) is the annihilation (creation) operator of the $k$-th photon mode, $g_k$ is its coupling strength to the emitter, and $\omega_k$ is the detuning from $\omega_0$ of mode $k$.  We consider pulses such that $\Omega_x(t) = \Omega_x =\Omega$ during the time $t_{\pi}$ of the $\pi$-pulses and zero otherwise. $\Delta(t)=\omega_1(t)-\omega_0$ is the time-dependent detuning of the TLS's transition frequency from the pulse carrier frequency.


We assume the system to be initially prepared in the excited state. In the absence of all control ($\Omega_x(t) = 0$ for all times), spontaneous decay will occur, and for a static detuning, the corresponding emission rate is $\Gamma = 2\pi\int g_k^2 \; \delta(\omega_k - \Delta) \; dk$;  We normalize our energy and time units so that this relaxation rate is $\Gamma = 2$, and the corresponding spontaneous emission line has a simple Lorentzian shape $1/(\omega^2 + 1)$, with half-width equal to $1$. By this process, all frequencies are measured in units of $\Gamma/2$. Previous studies have only studied the problem with static detuning and idealized instantaneous pulses~\cite{FotsoEtal_PRL2016, Fotso_JPhysB2019, LukinVuckovic_npj2020}. In the present treatment of the noisy quantum emitter, we assume the same relaxation rate $\Gamma$ as in the static detuning problem and focus on the time-dependence of $\Delta(t)$.

To study the dynamics of the TLS and evaluate its spectral properties, we analyze the time evolution of the emitter's density matrix operator:
\begin{eqnarray}
 \rho(t) &=& \rho_{ee}(t) |e\rangle \langle e| + \rho_{eg}(t) |e\rangle \langle g| \nonumber \\
 &+& \rho_{ge}(t) |g \rangle \langle e|  + \rho_{gg}(t) |g\rangle \langle g| \; ,
\label{eq:TrueDensMatr}
\end{eqnarray}
with the identities $\rho_{ge}^* = \rho_{eg}$, and $\rho_{ee} + \rho_{gg} = 1$. Using the approximation of independent rates of variation, the master equation governing the time-evolution of the density matrix is obtained by independently adding up, in the time-evolution of the matrix elements of $\rho$, terms due to the radiation bath, to the incident field, and the damping terms responsible for spontaneous emission.\cite{Cohen_Tannoudji_Book1992} 

For the model described above, the master equations characterizing the dynamics of the density matrix operator or optical Bloch equations in the rotating wave approximation can then be written as:

\begin{equation}
 \begin{bmatrix}
  \dot{\rho}_{ee} \\
  \dot{\rho}_{gg} \\  
  \dot{\rho}_{ge} \\
  \dot{\rho}_{eg} 
 \end{bmatrix}
=
\begin{bmatrix}
-\Gamma & 0 & -i\frac{\Omega_x(t)}{2} & i\frac{\Omega_x(t)}{2}\\
\Gamma & 0 &  i\frac{\Omega_x(t)}{2} & -i\frac{\Omega_x(t)}{2}\\
 -i\frac{\Omega_x(t)}{2} & i\frac{\Omega_x(t)}{2} & i\Delta -\frac{\Gamma}{2} & 0 \\
 i\frac{\Omega_x(t)}{2} & -i\frac{\Omega_x(t)}{2}  & 0 & -i\Delta -\frac{\Gamma}{2}
\end{bmatrix}
\begin{bmatrix}
  \rho_{ee} \\
  \rho_{gg} \\  
  \rho_{ge} \\
  \rho_{eg} 
\end{bmatrix}
\label{eq:masterEquation1}
\end{equation}

\section{Methods}
\label{sec:methods}

\subsection{Emission Spectrum}
\label{subsec:emission}
We will calculate the emission spectrum of the TLS that corresponds to the excitation probability of the detector in the narrow-band detector approach where the detector is modeled by a two-level absorber with a very sharp transition frequency.\cite{Scully_Zubairy_book1997} At a long time $T$, the emission spectrum can be expressed as:
\begin{eqnarray}
\label{eq:emissionEq1}
 P(\omega) &=& 2 A^2 \\
 & \times& \mathrm{Re} \left\{ \int_0^T dt \int_0^{T-t} d\theta \langle \sigma_+(t + \theta) \sigma_-(t) \rangle exp\left[ -i \omega \theta \right] \right\}. \nonumber 
\end{eqnarray}
Here, $A$ is a constant independent of the driving field parameters that does not affect the spectral shape but only affects the absolute scale of the spectrum. $\sigma_-(t)$ and  $\sigma_+(t+\theta)$ are the time-dependent operators of the TLS in the Heisenberg representation, and the angled brackets represent the expectation values that are taken with respect to the initial state.\\
To evaluate the two-time correlation function $\langle \sigma_+(t+\theta) \sigma_-(t) \rangle$, it is typically rewritten as a single-time expectation value~\cite{RF_Mollow_PhysRev1969, Scully_Zubairy_book1997, Loudon_book1983}:
\begin{equation}
 \langle \sigma_+(t+\theta) \sigma_-(t) \rangle  = \mathrm{Tr} \left[ \rho'(t + \theta) \sigma_+ \right]. 
 \label{eq:expVal_sigmaP_sigmaM}
\end{equation}   
 Here, $\sigma_+$ and $\sigma_-$ are the time-independent operators in
the Schr\"odinger picture and $\rho^{\prime}(t+\theta)$ is obtained from the original density matrix operator by $\rho^{\prime}(t) = \sigma_- \rho(t)$ at time $t$ and then evolved under the same master equations (\ref{eq:masterEquation1}) from time $t$ up to time $t+\theta$. 

Taking advantage of expression (\ref{eq:expVal_sigmaP_sigmaM}), the emission spectrum is calculated numerically using the following recipe. The time axis is discretized into equal time slices of width $\Delta t = \tau/N_t$. Where $N_t$ is the number of time slices in a pulse interval of width $\tau$. Starting at time $t=0$ where the initial conditions are known ($\rho_{ee} = 1,\; \rho_{gg} = 0,\; \rho_{eg} = 0,\; \rho_{ge} = 0$), we integrate the master equation to obtain the matrix elements $\rho_{ee},\; \rho_{eg},\; \rho_{ge},\; \rho_{gg}$ from $t$ to $t + \Delta t$, first for a freely evolving TLS with Rabbi frequency $\Omega_x(t) = 0$ until time $\tau - t_{\pi}$, then from time $\tau - t_{\pi}$ to time $\tau$, in the presence of the driving field ($\Omega_x(t) = \Omega$). The matrix elements of $\rho$ at the end of this pulse interval are then used as initial values for the next pulse interval and the process is repeated for a number $N_p$ of consecutive pulse intervals, resulting in the knowledge of $\rho(t)$ and thus $\rho^{\prime}(t)$ for $t \in [0, T]$ with $T = N_p \tau$. This is then followed of by the integration of the master equation starting from each time $t \in [0, T]$ to produce $\rho^{\prime}(t + \theta)$ for $\theta \in [0, T - t]$. From this, we obtain the the correlation function $\langle \sigma_+(t+\theta) \sigma_-(t) \rangle$. Finally, we perform the Fourier transform with respect to $\theta$ and the integration over $t$ to obtain $P(\omega)$ which is our emission spectrum.  Throughout this integration process, we use a time-dependent detuning that is obtained by generating a random Gaussian distribution with mean value $\Delta_0$, standard deviation $\sigma_{\Delta}$ and a correlation time $\tau_c$.

\subsection{Two-Photon Interference}
\label{subsec:TPI}

After studying the emission spectrum, we next consider a HOM-type two-photon interference operation between two distant quantum emitters, each in its own noisy environment resulting in different inhomogeneously broadened spectral signatures. Photons from emitter $E_1$ with mean detuning $\Delta_{01}$ and standard deviation $\sigma_{\Delta 1}$ and from emitter $E_2$ with mean detuning $\Delta_{02}$ and standard deviation $\sigma_{\Delta 2}$, at spacetime locations $1$ and $2$ respectively, are sent to the input ports of a $50:50$ beam splitter and then measured at detectors $D_1$ and $D_2$ at spacetime locations $3$ and $4$ beyond the output ports of the beam splitter. The emitters can each be independently modeled by the Hamiltonian (\ref{eq:hamiltonian_1}). We want to evaluate the second order coherence equivalent to the intensity correlation at the detectors $D_1$ and $D_2$ in the presence of the control protocols made of identical finite width pulses driving the respective emitters $E_1$ and $E_2$. This intensity correlation is:
\begin{equation}
\label{eq:G2_34_def}
 G^{(2)}_{34}(t,\theta) = \langle a_3^{\dagger}(t) a_4^{\dagger}(t+\theta)a_4(t+\theta) a_3(t) \rangle.
\end{equation}
$a_i^{\dagger}(t)$ ($a_i(t)$) is the creation (destruction) operator of a photon  at detector $i$. From (\ref{eq:G2_34_def}), we extract the intensity correlation corresponding to the measured cross-correlation in the Hanburry Brown and Twiss setup\cite{HanburryBrownTwiss_1956, KirazAtature_PRA_2004}:
\begin{equation}
\label{eq:g2_34_integrated}
 g^{(2)}_{34}(\theta) = \lim_{T \to \infty} \int_0^T G_{34}^{(2)}(t, \theta) \; dt.
\end{equation}
This correlation function is rewritten in terms of the two-time correlation functions at the emitters denoted by the $i$ index, $g_i(t, \theta) = \langle \sigma_{+ i}(t) \sigma_{- i}(t + \theta) \rangle$, that can then be expressed as single-time expectation values similarly to (\ref{eq:expVal_sigmaP_sigmaM})~\cite{RF_Mollow_PhysRev1969, Scully_Zubairy_book1997, Loudon_book1983, Scully_Zubairy_book1997, Fotso_JPhysB2019, Fotso_TPI_PRB_2019}.
To evaluate $g_i(t, \theta)$, we can then use the same procedure employed in Ref.\cite{Fotso_TPI_PRB_2019} where the master equation (\ref{eq:masterEquation1}) is integrated on the discretized time axis following steps similar to those highlighted above for the emission spectrum\cite{FotsoDobrovitski_Absorption, Fotso_JPhysB2019}.

\subsection{Ensemble of Quantum Emitters}
\label{subsec:ensembleOfQEs}

Finally, we consider the case of an ensemble of two-level systems with each emitter sitting in its own specific environment that sets its emission frequency to be independent of other emitters' in the ensemble. Here, we consider the case where the ensemble has emitters with static detunings distributed randomly according to a Gaussian distribution. The ensemble is assumed to be dilute so that we can neglect dipole-dipole interactions between member emitters. In this situation, often encountered in spectroscopy experiments, the emission frequency of the ensemble can be obtained by adding up contributions from individual emitters. For a typical ensemble, this emission spectrum may become broadly spread out without clearly identifiable spectral  features\cite{RF_Heitler_Book1960}. This situation can be illustrated in a manner similar to that of FIG.\ref{fig:schematic1} where instead of a time axis to track the individual emitter, with snapshots at different times, one observes at a given moment multiple quantum emitters in the ensemble. We consider this system when it is placed under the influence of our finite-width periodic pulse sequence and we aim to assess the effect on the emission spectrum of the ensemble. Each emitter in the ensemble can be described by the Hamiltonian (\ref{eq:hamiltonian_1}) with its detuning $\Delta$ with respect to the pulse carrier frequency. The contributions to the emission spectrum of the individual emitters are obtained following the procedure highlighted in section (\ref{subsec:emission}).

\begin{figure}[htbp] 
\begin{center}
\includegraphics[width=7.0cm]{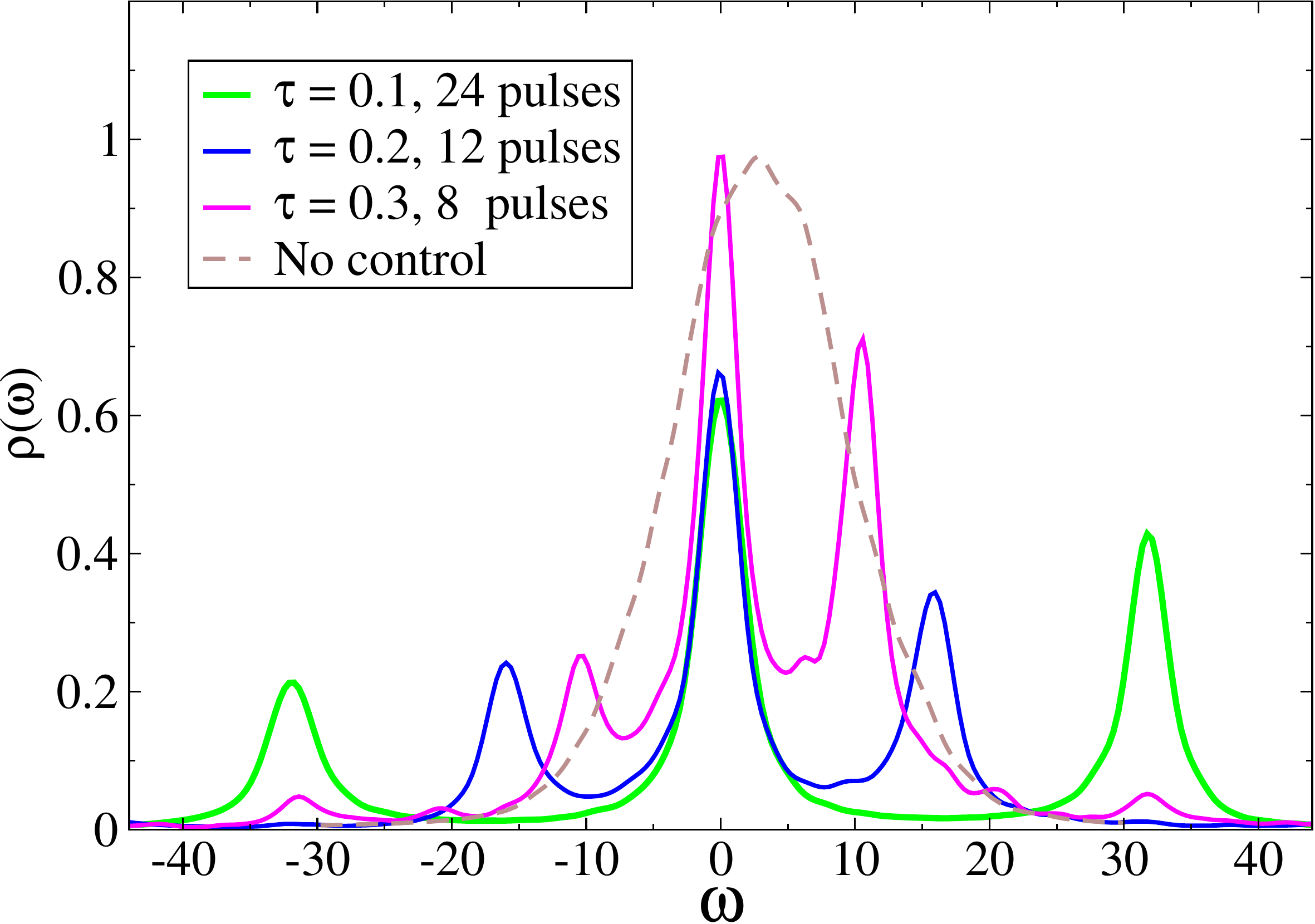}
    \caption{Emission spectrum for the fluctuating quantum emitter following a Gaussian random distribution of $\Delta(t)$ with $\Delta_0 = 3.0$ for $\tau = 0.1$ (green), $\tau = 0.2$ (blue), $\tau = 0.3$ (magenta) after a total time $t=2.4$ in all cases with Rabbi frequency $\Omega=35$. The dashed brown line shows the emission spectrum of the system in the absence of the control pulse sequence.} 
\label{fig:emissionSpectFixedDelta_3Taus}
\end{center}
\end{figure}

\begin{figure}[htbp] 
\begin{center}
\includegraphics[width=8.0cm]{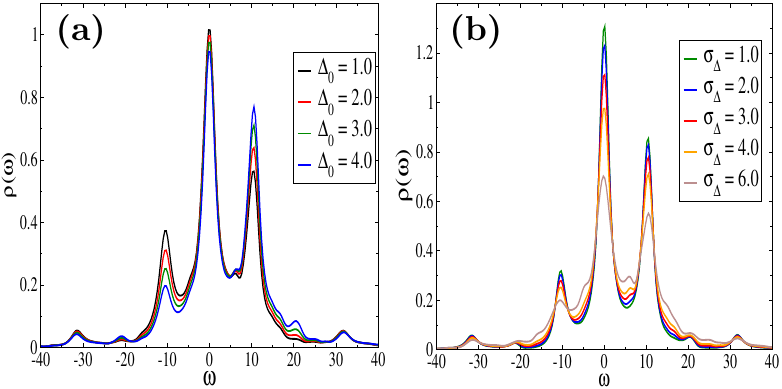}
\caption{ Emission spectrum of the noisy TLS under a periodic pulse sequence with inter-pulse delay $\tau = 0.3$ and Rabbi frequency $\Omega= 35.$ (a) For fixed variance ($\sigma_{\Delta} = 4.0$ ) with average detuning values $\Delta_0 = 1.0$ (black), $\Delta_0 = 2.0$ (red), $\Delta_0 = 3.0$ (green), $\Delta_0 = 4.0$ (blue). (b) For fixed average detuning value $\Delta_0 = 3.0$ with standard deviation $\sigma_{\Delta}=1.0$ (green), $\sigma_{\Delta}=2.0$ (blue), $\sigma_{\Delta}=3.0$ (red), $\sigma_{\Delta}=4.0$ (orange), $\sigma_{\Delta}=6.0$ (brown).} 
\label{fig:emissionSpectFixedTau2X1}
\end{center}
\end{figure}

\section{Results}
\label{sec:results}

\subsection{Emission Spectrum}
We consider random Gaussian distributions with a correlation time that is of the order of the inter-pulse delay or less. We find that our emission spectra have little dependence on this parameter and so in what follows our results are presented for $\tau_c \sim 0.03$. FIG.\ref{fig:emissionSpectFixedDelta_3Taus} presents the emission spectrum for a TLS with $\Delta(t)$ such that the average value is $\Delta_0 = 3.0$ and the standard deviation is $\sigma_{\Delta} = 4.0$. The solid green line corresponds to inter-pulse time delay of $\tau = 0.1$, the blue line to $\tau = 0.2$, and the pink line to $\tau = 0.3$. The Rabbi frequency is $\Omega = 35$ for all pulse sequences.  

The dashed brown line shows the emission spectrum of this system measured over the same duration when the system is not subject to any control field. Clearly, the pulse sequence produces on this noisy system, a spectrum similar to that reported in the case of a static detuning. The protocol maintains nearly $50\%$ of the spectral weight at the pulse carrier frequency. The controlled spectrum also features satellite peaks at integer multiples of $\pm \pi/\tau$ with spectral weights suppressed away from the central peak. The bulk of the emission spectrum is refocused to the pulse carrier frequency even for fairly broad pulses.

Next, we examine this emission spectrum as a function of random distributions of detunings (as a function of the mean value and the standard deviation of the distribution or the typical width of the distribution). FIG.\ref{fig:emissionSpectFixedTau2X1}-(a) shows the emission spectrum under the same pulse sequence of period $\tau=0.3$ and Rabbi frequency $\Omega=35$ for the same standard deviation $\sigma_{\Delta}=4.0 $ with average detuning value $\Delta_0 = 1.0$ (black line), $\Delta_0 = 2.0$ (red line), $\Delta_0 = 3.0$ (green line), $\Delta_0 = 4.0$ (blue line). FIG.\ref{fig:emissionSpectFixedTau2X1}-(b) shows the emission spectrum under the same pulse sequence (inter-pulse delay $\tau=0.3$ and Rabbi frequency $\Omega=35$), for fixed average detuning value $\Delta_0 = 3.0$ with standard deviation $\sigma_{\Delta}=1.0$ (green line), $\sigma_{\Delta}=2.0$ (blue line), $\sigma_{\Delta}=3.0$ (red line), $\sigma_{\Delta}=4.0$ (orange line), $\sigma_{\Delta}=6.0$ (brown line). The lineshape is overall preserved for a broad range of parameters. For the narrower distribution and for the smaller average detuning, the central peak contains more of the spectral weight. The refocusing of the spectral peak is deteriorated for broader distributions and for large average detuning values. However, we observe that overall, the protocol remains effective as long as the distribution of detunings is such that $ \Delta \times \tau \lesssim 1$ for most of the detuning values.

\begin{figure}[htbp] 
\begin{center}
\includegraphics[width=8.50cm]{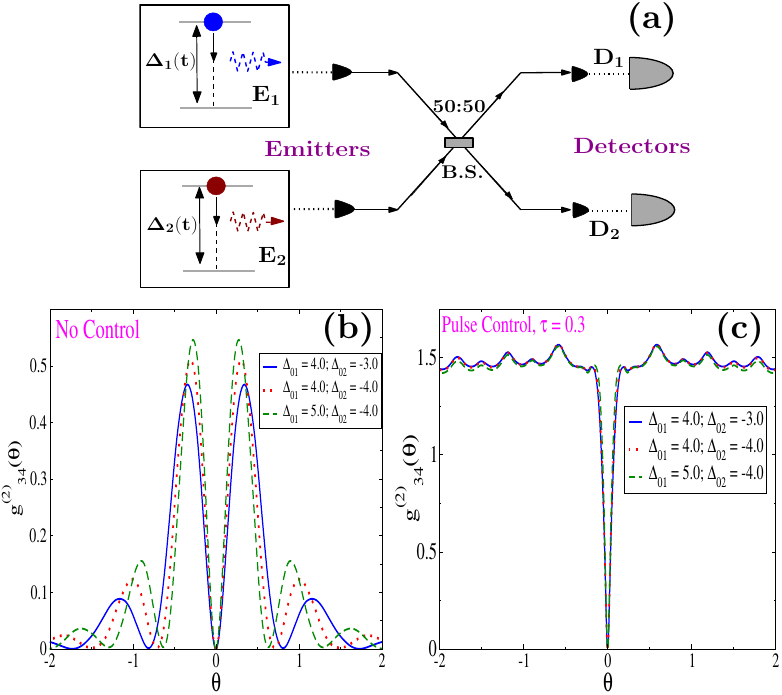}
    \caption{(a) Schematic representation of the two-photon interference operation between two distant emitters at spacetime locations $1$ and $2$ with fluctuating emission frequencies characterized by $E_1(t)$ and $E_2(t)$. Photons from the emitters are sent to a 50:50 beam splitter and then measured at detectors $D_1$ and $D_2$ at locations $3$ and $4$. Cross-correlation function at the detectors as a function of the delay time for two emitters with Gaussian random detunings with average values $\Delta_{01} = 4.0$ and $\Delta_{02} =-3.0 $ (blue solid line), $\Delta_{01} = 4.0$ and $\Delta_{02} =-4.0 $  (red dotted line), $\Delta_{01} = 5.0$ and $\Delta_{02} =-4.0 $   (green dashed line)  in the absence of any control protocol (b) and under a periodic pulse sequence with inter-pulse delay $\tau = 0.3$ (c). The standard deviation is $\sigma_{\Delta 1} = \sigma_{\Delta 2} = 6.0$ in all cases.} 
\label{fig:tpi2X1}
\end{center}
\end{figure}

\subsection{Two-Photon Interference}

After studying the control of the emission spectrum of the isolated  quantum emitter in a noisy environment, we consider the two-photon interference operation between two such emitters as depicted schematically in FIG.\ref{fig:tpi2X1}-(a). The cross-correlation function at the detectors $g^{(2)}_{34}(\theta)$ as a function of delay time $\theta$ is shown in FIG.\ref{fig:tpi2X1}-(b) and FIG.\ref{fig:tpi2X1}-(c) respectively without control protocols and when the two emitters are driven by a pulse sequence of period $\tau = 0.3$ with Rabbi frequency $\Omega = 35$. The figures show $ g^{(2)}_{34}(\theta)$ as a function of $\theta$ for pairs of emitters with average detuning values of $\Delta_{01} = 4.0$ and $\Delta_{02} = 3.0 $ (blue solid line), $\Delta_{01} = 4.0$ and $\Delta_{02} = -4.0$ (red dotted line), $\Delta_{01} = 5.0$ and $\Delta_{02} = -4.0$ (green dashed line). Note that $g^{(2)}_{34}(\theta)$ vanishes in both cases at $\theta=0$. However, in the absence of the control protocol it also vanishes periodically at times that are integer multiples of $\sim \pi/( \Delta_{01} -\Delta_{02})$ and overall decreases in magnitude with increasing delay times, while it stays finite for all other times in the presence of the control protocol including for emitters in significantly different environments $\Delta_{01} -\Delta_{02} \sim 10 \Gamma $. Where $\Gamma$ is the free emission lifetime of an individual emitter.

\begin{figure}[htbp] 
\begin{center}
\includegraphics*[width=8.0cm]{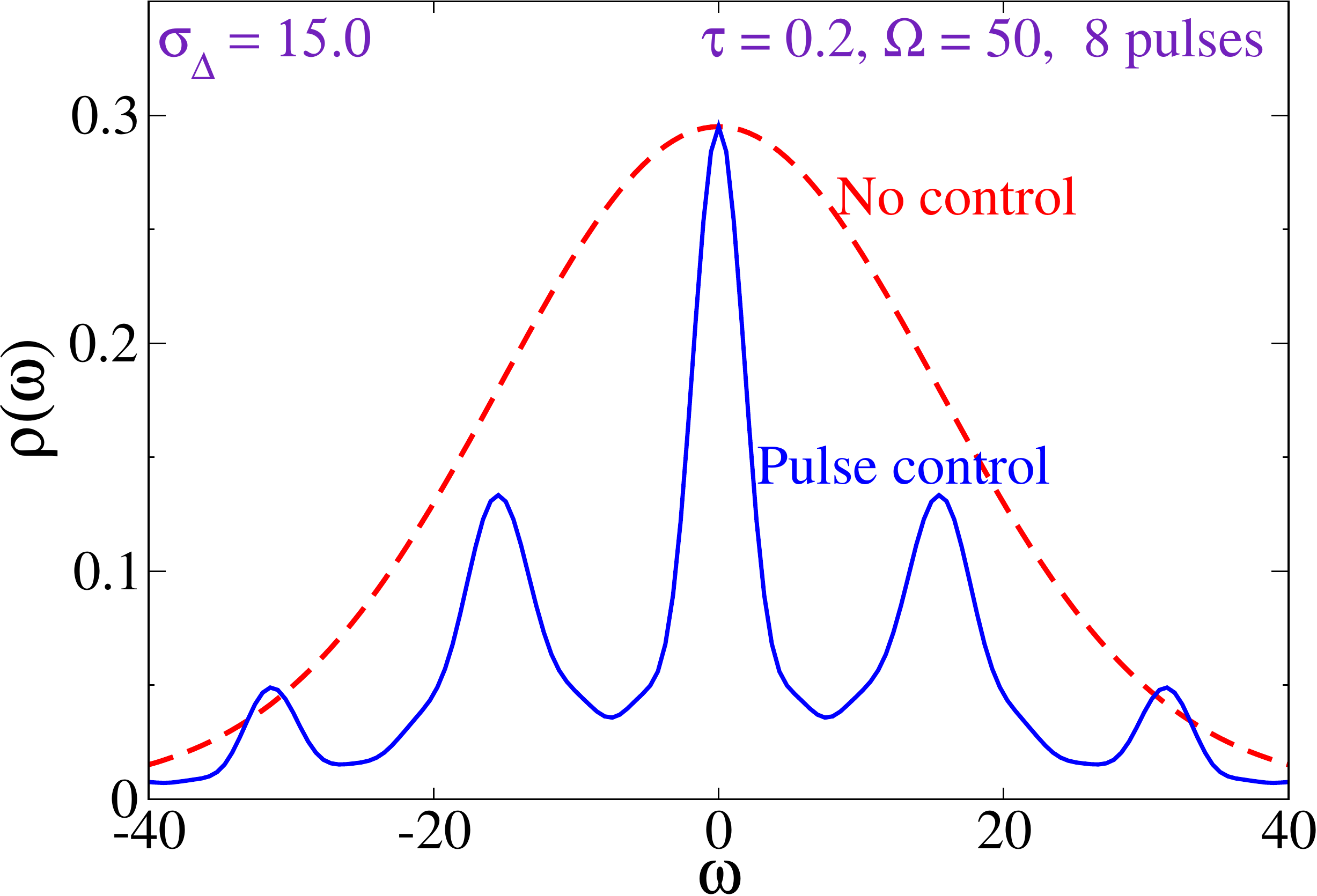}
\caption{Emission spectrum of an ensemble of quantum emitters with the emission frequencies of individual emitters spread across a random Gaussian distribution of detunings with standard deviation $\sigma_{\Delta}=15.0$ and average value $\Delta_{ave}=0$. The dashed red line shows the emission spectrum of the system in the absence of any control protocol while the blue solid line shows the emission spectrum when the system is under the influence of a periodic sequence of $\pi$ pulses with inter-pulse delay $\tau =0.2$ and with Rabbi frequency $\Omega = 50$ after 8 pulses. The spectrum with no control has been rescaled to have the same maximum as the spectrum under the pulse sequence.} 
\label{fig:ensembleInhomroad}
\end{center}
\end{figure}

\subsection{Ensemble of Two-Level Systems}

We now consider a dilute ensemble of quantum emitters described by two-level systems. An emitter $i$ in the ensemble has its own particular environment and its detuning is set to a value $\Delta_i$. The $\Delta_i$'s follow a random Gaussian distribution with average value $\Delta_{ave} = 0$ and a standard deviation $\sigma_{\Delta} = 15.0$. FIG.\ref{fig:ensembleInhomroad} shows the emission spectrum of this ensemble when it is placed under the influence of a pulse sequence  of period $\tau=0.2$ with Rabbi frequency $\Omega=50$. The spectrum is calculated for 8 pulses but the general lineshape is established after 2 to 4 pulses and further time mostly results in larger peak amplitudes. Note that the amplitude of the spectrum in the absence of the control protocol is rescaled to match that of the spectrum in the presence of the control fields. Overall, the emission spectrum that has a broad Gaussian lineshape in the absence of the control protocol is refocused by the control protocol to result in a lineshape with a central peak at the pulse carrier frequency ($\omega=0$ in the rotating frame) that has the linewidth $\Gamma$ of an individual isolated emitter flanked by satellite peaks at integer multiples of $\pm \pi/\tau$.

\section{conclusion}
\label{sec:conclusion}
We have examined spectral properties of quantum emitters in noisy environments manifested by a random Gaussian distribution of detunings, as a function of time for individual quantum emitters and by a random distribution of detunings across an ensemble of two-level systems. For individual emitters, we characterized the emission spectrum under the effect of a periodic sequence of finite-width pulses. Our results indicate that for a broad range of parameters, the emission spectrum of a noisy quantum emitter can be well controlled by the pulse sequence. When two different such noisy individual two-level systems are used in a HOM-type two-photon interference, we find that the periodic sequence of finite width pulses effectively restores two-photon indistinguishability between two the two spectrally different systems. Finally, for an ensemble of quantum emitters, with individual emitters that have randomly distributed emission frequencies so that the ensemble would produce an inhomogeneously broadened emission spectrum, we show that the control protocol can refocus the emission spectrum to a lineshape with a central peak that has the linewidth of an individual quantum emitters. These results demonstrate for a model of spectrally noisy two-level system the ability to optimize spectral properties with an external control field in the form of a pulse sequence.\\

\section*{Acknowledgments} 
This work is supported by the National Science Foundation under Grant No. PHY-2014023. We thank Tim Schr\"oder for useful conversations.

\end{document}